\documentclass[aps,twocolumn,showpacs,preprintnumbers,amsmath,amssymb,showpacs,showkeys]{revtex4}
\usepackage{amssymb}
\usepackage{amsmath}
\usepackage{amsfonts}
\usepackage{epsfig}
\usepackage{graphicx}
\usepackage{tabularx}
\newcommand{\seq}{\begin{subequations}}
\newcommand{\sen}{\end{subequations}}
\newcommand{\eq}{\begin{eqnarray}}
\newcommand{\en}{\end{eqnarray}}

\def\shiftdown#1{#1\llap{\lower.04ex\hbox{#1}}}

\def\arraystretch{1.5}

\begin{document}

\title{Heavy quarkonium properties from Cornell potential using variational method and supersymmetric quantum mechanic}

\author{Alfredo Vega$^{1,2}$ and
        Jorge Flores$^1$
\vspace*{1.2\baselineskip}\\
}

\affiliation{
$^1$ Instituto de F\'isica y Astronom\'ia,
     Universidad de Valpara\'iso,\\
     Avenida Gran Breta\~na 1111, Valpara\'iso, Chile
\vspace*{1.0\baselineskip} \\
$^2$ Centro de Astrof\'isica de Valpara\'iso, Universidad de Valpara\'iso,\\
Avenida Gran Breta\~na 1111, Valpara\'iso, Chile
\vspace*{1.2\baselineskip} \\
}

\date{\today}

\begin{abstract}

Using the variational method and supersymmetric quantum mechanic we calculate in a approximate way eigenvalues, eigenfunctions and wave functions at origin of Cornell potential. We compare results with numerical solutions for heavy quarkonia $c\bar{c}$, $b\bar{b}$ y $b\bar{c}$.

\end{abstract}



\maketitle

\section{Introduction}

Since the discovery of  $J/\psi$ in 1974 \citep{Aubert:1974js,Augustin:1974xw} the study of heavy quarkonium has been very valuable in hadron physics, because they involve non perturbative aspect of QCD and there are many experimental data involving those hadrons \cite{Voloshin:2007dx,Patrignani:2012an,Brambilla:2010cs}. From a theoretical standpoint, heavy quarkonium has been studied from several approaches \cite{Grinstein:1998xb,Lucha:1991vn}, we can stand out for his simplicity, and because it corresponds to a successful approach, the non relativistic potential models, where quark interaction is modelled using a potential energy in the usual Schr\"odinger equation. The literature about quark potentials is huge, so here we add a small number of references \cite{Lucha:1991vn,Eichten:1978tg,Eichten:1979ms,Richardson:1978bt,Buchmuller:1980su,Gupta:1981pd,Gupta:1986xt,Gonzalez:2003gx,Dib:2012vw}, although incomplete is representative and is a good starting point to introduce on this topic. One of the first potential proposed was the Cornell potential \cite{Eichten:1978tg,Eichten:1979ms}, that corresponds to a coulombian potential plus a linear confinement term. In this way Cornell potential considers general properties of quark interactions. 

Schr\"odinger equation with Cornell potential do not have analytical solutions, and although it is possible to solve it using numerical methods \cite{Lucha:1998xc,DomenechGarret:2008sr}, it is always interesting to obtain an approximate analytical solutions that offer a possibility of additional discussions.

In this work, we solve in an approximate way the Schr\"odinger equation with Cornell potential using a procedure that correspond to an adaptation method suggested in \cite{Gozzi:1993ea,CooperVar}, that considers usual variational methods with supersymmetric quantum mechanic (SUSY QM). Some additional examples using SUSY QM and variational method can be found in \cite{DrigoFilho:1995fn,DrigoFilho:1999aw,PeglowBorges:2001hs}.

The use of  SUSY QM \cite{Cooper:1994eh}, born at the beginning of the eighties in studies of supersymmetry breaking in quantum field theories with extra dimensions \cite{Witten:1981nf}, allows to get isospectral potentials to the original potential, with the particularity that the ground state of the original potential is not present in the spectrum of isospectral associated potential, so the ground state of the supersymmetric partner potential is related with the first excited state of original potential.This procedure can be repeated in order to get successive potentials whose ground states are related by some standard transformations in SUSY QM, with the different states of the original potential, Cornell in our case. So in this way, if we use the variational method to get solutions for the ground state of different supersymmetric partners of the Cornell potential we can obtain the spectrum and wave functions for heavy quarkonium. Notice that standard variational method has been used to study heavy quarkonium properties considering different phenomenological quark potentials \cite{Ding:1998qr,Boroun}.

The procedure described in the previous paragraph is used in this work to get approximated eigenvalues and eigenfunctions for the Schr\"odinger equation with Cornell potential, and we are using it to study heavy quarkonium $c\bar{c}$, $b\bar{b}$ and $b\bar{c}$, paying special attention to the wave function at the origin (WFO), an important quantity that it is involved in calculations of heavy quarkonium decay rates.

This paper is structured as follow. In section II we sumarize the main ingredient of SUSY QM used in this work. Section III it is dedicated to approximated calculations of energies, wave functions and WFO for heavy quarkonium using variational method and SUSY QM and in section IV we discuss our results and conclusions.

\section{Basics of SUSY QM}

In this section we summarize the main ingredients of SUSY QM that we will use in the next sections, in order to calculate heavy mesons properties using Cornell potential. For more details we suggest Ref. \cite{Cooper:1994eh}.

Let us consider Schr\"odinger equation for the ground state with eigenvalue equal to zero, so the wave function $\psi_{0}$ obeys
\begin{equation}
\label{H1}
H_{1} \psi_{0}(x) = - \frac{\hbar^{2}}{2 m} \frac{d^{2} \psi_{0}(x)}{dx^{2}} + V_{1}(x) \psi_{0}(x) = 0
\end{equation}
then
\begin{equation}
V_{1}(x) = \frac{\hbar^{2}}{2 m} \frac{\psi_{0}^{''}(x)}{\psi_{0}(x)}.
\end{equation}

The hamiltonian $H_{1}$ can be factorized as
\[
H_{1} = A^{\dagger} A,
\]
where
\[
A = \frac{\hbar}{\sqrt{2 m}} \frac{d}{dx} + W(x)
~~~~y~~~~
A^{\dagger} = - \frac{\hbar}{\sqrt{2 m}} \frac{d}{dx} + W(x).
\]

With this we can see that for a known $V_{1}$, the superpotential $W$ satisfies a Riccati equation 
\[
V_{1}(x) = - \frac{\hbar}{\sqrt{2 m}} \frac{d W(x)}{dx} + W^{2}(x).
\]

The solution for $W(x)$ in terms of the ground state wave function is
\begin{equation}
\label{W}
W(x) = -\frac{\hbar}{\sqrt{2 m}} \frac{\psi^{'}_{0}(x)}{\psi_{0}(x)}.
\end{equation}

With operators $A$ and $A^{\dagger}$ it is possible to built a new hamiltonian $H_{2}$ given by
\[
H_{2} = A A^{\dagger},
\]
that can be expressed by
\[
H_{2} = - \frac{\hbar^{2}}{2 m} \frac{d^{2}}{dx^{2}} + V_{2}(x),
\]
where
\[
V_{2}(x) = \frac{\hbar}{\sqrt{2 m}} \frac{d  W(x)}{dx} + W^{2}(x).
\]

Potentials $V_{1}(x)$ and $V_{2}(x)$ are known as supersymmetric partner potentials, and they have several interesting properties (see \cite{Cooper:1994eh}).

Eigenvalues and eigenfunctions of $H_{1}$ and $H_{2}$ are related by
\begin{equation}
E_{n}^{(2)} = E_{n+1}^{(1)}~~~~~;~~~~~E_{0}^{(1)} = 0
\end{equation}
\begin{equation}
\psi_{n}^{(2)} = \frac{1}{\sqrt{E_{n+1}^{(1)}}} A \psi_{n+1}^{(1)}
\end{equation}
and
\begin{equation}
\label{WFtransformada}
\psi_{n+1}^{(1)} = \frac{1}{\sqrt{E_{n}^{(2)}}} A^{\dagger} \psi_{n}^{(2)}.
\end{equation}

We want to pay special attention to the relationship in the spectrum of $H_{1}$ and $H_{2}$, because with exception of the ground state of $H_{1}$, that does not appear in $H_{2}$, the other levels are shared in both hamiltonians, i.e except $E_{0}^{(1)}$, potentials $V_{1}$ and $V_{2}$ are isospectrals. Successives applications of this procedure led us to obtain a family of isospectral potentials where, as you can see in FIG. 1, the ground state of $V_{2}$ is related with the first excitated level of $V_{1}$, ground state of $V_{3}$ is related with the first excitated level of $V_{2}$ and the second level of $V_{1}$, and so on to get more levels of the original potential $V_{1}$.

\begin{figure}[t]
\center
\includegraphics[width=8.0cm]{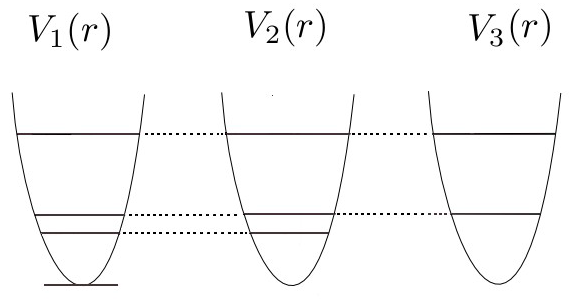}
\caption{This figure show a schematic representation of potential $V_{1}$ and his supersymmetric partners $V_{2}$ and $V_{3}$ with his corresponding spectrum, and show that the ground state of $V_{1}$ is not present in $V_{2}$, and the ground state of $V_{2}$ is not present in $V_{3}$, and so on. Strictly speaking the shape of each potential is different, but we disregard this in order to present that the ground state of one potential is not present in his supersymmetric partner.}
\end{figure}

According to the previous paragraph the ground state of $H_{2}$ correspond to the first excited level of $H_{1}$. This fact is specially interesting for us, because usual variational method are good tools to get approximated values for the ground states in the Schr\"odinger equation, so we can use this simple procedure to get approximated solutions for different supersymmetric partner  potentials, and in this way to get solutions for different levels in a potential of interest, as the Cornell potential.

\section{Solutions for Cornell potential with variational methods and SUSY QM}

In this section we use the procedure described in the previous sections, suggested in \cite{Gozzi:1993ea,CooperVar}, and we adapt it to the study of heavy quarkonium properties using the Cornell potential. We calculate energies, wave functions and WFO for S states for $c\bar{c}$, $b\bar{b}$ and $b\bar{c}$.

We consider $U_{nl}(r) = r R_{nl} (r)$, that by considering $\hbar = 1$ satisfies

\begin{equation}
\label{Sch}
- \frac{1}{2 \mu} \frac{d^{2} U_{nl}(r)}{dr^{2}} + V_{ef}(r) U_{nl}(r) = E_{n} U_{nl} (r),
\end{equation}
where $\mu$ the reduced mass for heavy quarkonium considered, and
\[
V_{eff}(r) = V(r) + \frac{l(l+1)}{2 \mu r^{2}}.
\]

In this work we consider $l=0$ (S states), thus the effective potential is equal to the Cornell potential,

\[
V_{eff}(r) = V(r) = - \frac{\kappa}{r} + \beta r,
\]
where the parameters involved are the same used in the usual calculations that consider this potential, i.e
\[
\kappa = 0.52~~~~;~~~~\beta = \frac{1}{(2.34)^{2}} GeV^{2},
\]
and the quark masses are
\[
\mu_{c} = 1.84 \biggl[\frac{GeV}{c^{2}}\biggr]~~~~y~~~~\mu_{b} = 5.18 \biggl[\frac{GeV}{c^{2}}\biggr].
\]

For the variational method we consider the trial wave function
\begin{equation}
\label{FnEnsayo}
U(r) = N r^{\gamma} e^{-a r^{b}},
\end{equation}
that will be used to get the ground state in the Cornell potential and in his supersymmetric partners used to calculate the successive levels. In this wave function $\gamma$ take the values 1, 2, 3, ... depending on if we are calculating the groung state of potentials $V_{1}$, $V_{2}$, ... (according section II), changing this parameter in this way is important, because in order to get approximations for the wave functions of the Cornell potential for different levels, it will be necessary to apply successive transformations defined by (\ref{WFtransformada}) and this choice turns out to be the only possibility to get finite WFO with a trial wave function as (\ref{FnEnsayo}). Parameters $a$ and $b$ are variational parameters and $N$ is the normalization constant, given by
\[
N={\sqrt{\frac{\left( 2a\right) ^{\frac{1+2\gamma}{b}}b}{\Gamma \left( \frac{1+2\gamma}{%
b}\right) }}}.
\]

To calculate the ground state for the Cornell potential we use (\ref{Sch}) and the trial wave function with $\gamma =1$, so the expectation value for the energy
\[
E = -\frac{1}{2 \mu} \int\limits_{0}^{\infty} U(r) \frac{d^{2}}{dr^{2}} U(r) dr + \int\limits_{0}^{\infty} \biggl( - \frac{\kappa}{r} + \beta r \biggr) U^{2}(r) dr.
\]

By using a $U(r)$ given by (\ref{FnEnsayo}), we get an expectation value of the energy that depends on parameters $a$ and $b$ ($E(a,b)$), and by minimizing this, we found an approximated value for the energy and we get values for the parameters $a$ and $b$. In this case we will call $a_{0}$ y $b_{0}$ for parameters associated to the ground state.

\begin{table}
\begin{center}
\caption{Energy values (in GeV) heavy quarkonium $c\bar{c}$, $b\bar{b}$ and $b\bar{c}$. column ``Exact'' solution correspond to numerical calculations using mathschroe.nb with step $h=0.00001$, and column ``Ours'' shows energies calculated in this work.}
\vspace*{.25cm}

\def\arraystretch{1.5}
\begin{tabular}{|c|c|c|c|c|c|c|}
\hline
~~ & \multicolumn{2}{|c|}{$c\overline{c}$} & \multicolumn{2}{|c|}{$b%
\overline{b}$} & \multicolumn{2}{|c|}{$b\overline{c}$} \\ \hline
$E_{n}$ & $Exact$ & $Ours$ & $Exact$ & $Ours$ & $Exact$ & $Ours$
\\ \hline
$1s$ & $0.2575$ & $0.2578$ & $-0.1704$ & $-0.1702$ & $0.1110$ & $0.1113$ \\ 
\hline
$2s$ & $0.8482$ & $0.8096$ & $0.4214$ & $0.3579$ & $0.6813$ & $0.6324$ \\ 
\hline
$3s$ & $1.2720$ & $1.1427$ & $0.7665$ & $0.5612$ & $1.0686$ & $0.9065$ \\ 
\hline
\end{tabular}

\end{center}
\end{table}

\begin{center}
\begin{figure*}
  \begin{tabular}{c c c}
    \includegraphics[width=2.3 in]{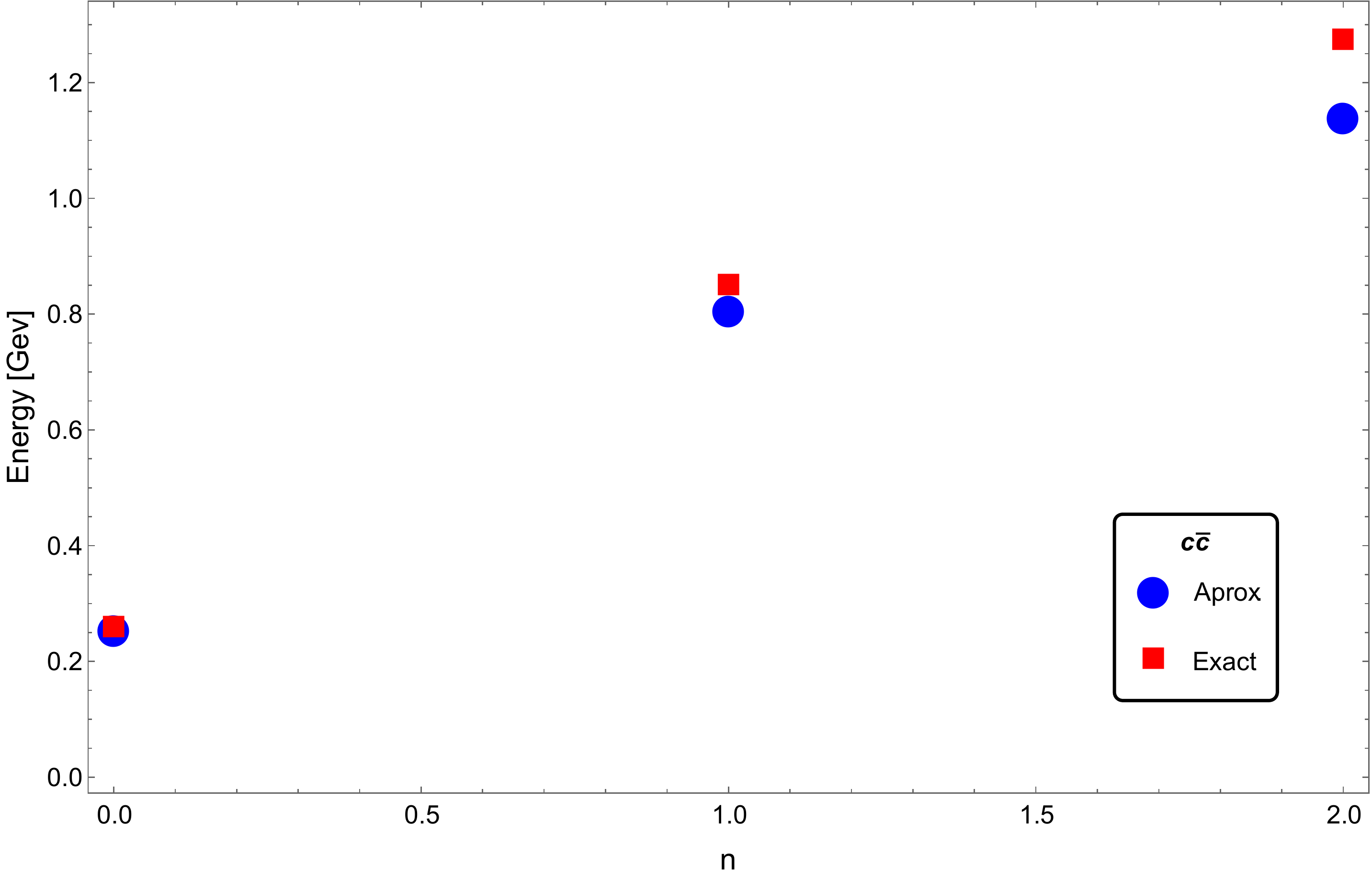}
    \includegraphics[width=2.3 in]{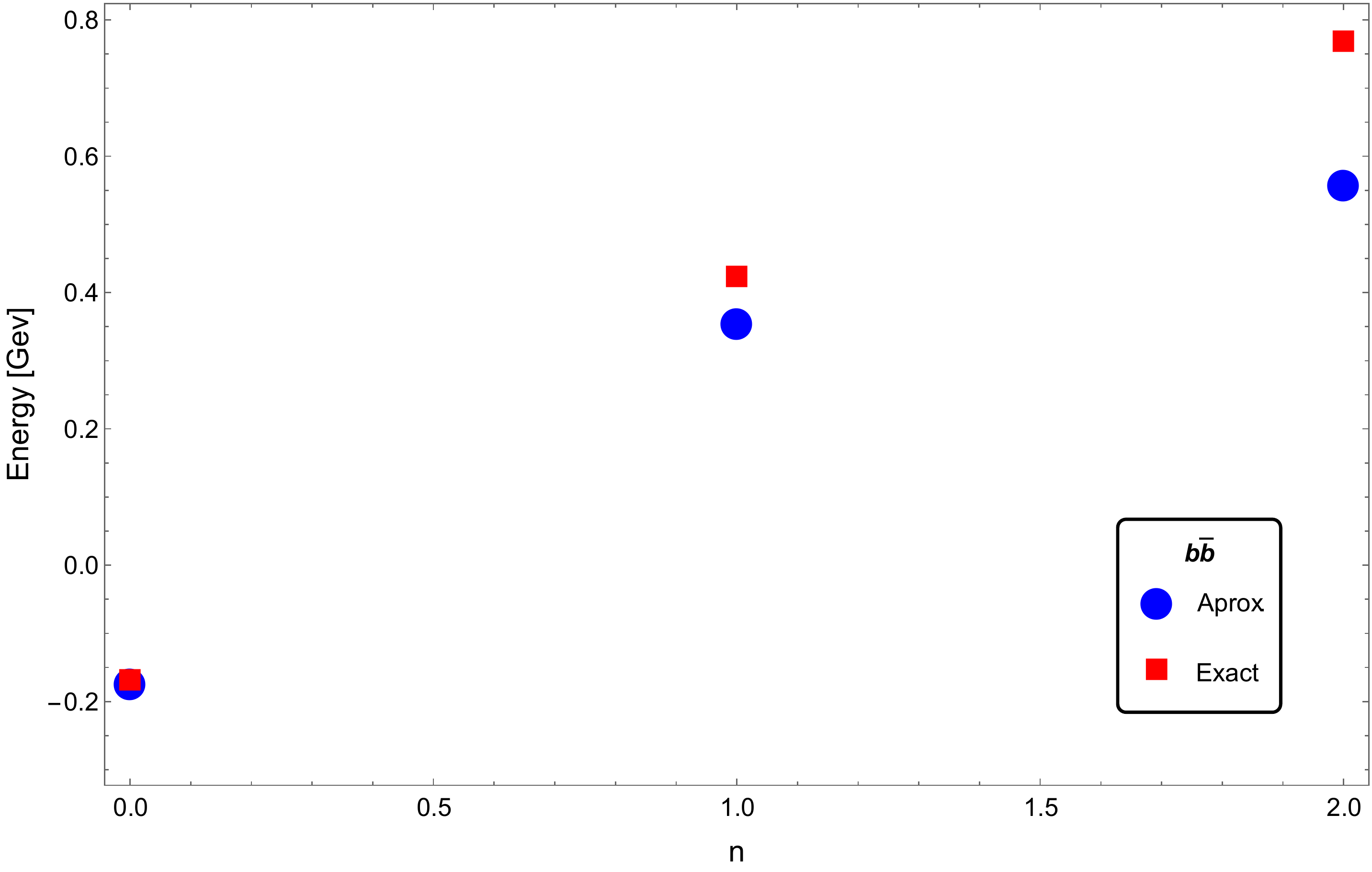}
    \includegraphics[width=2.3 in]{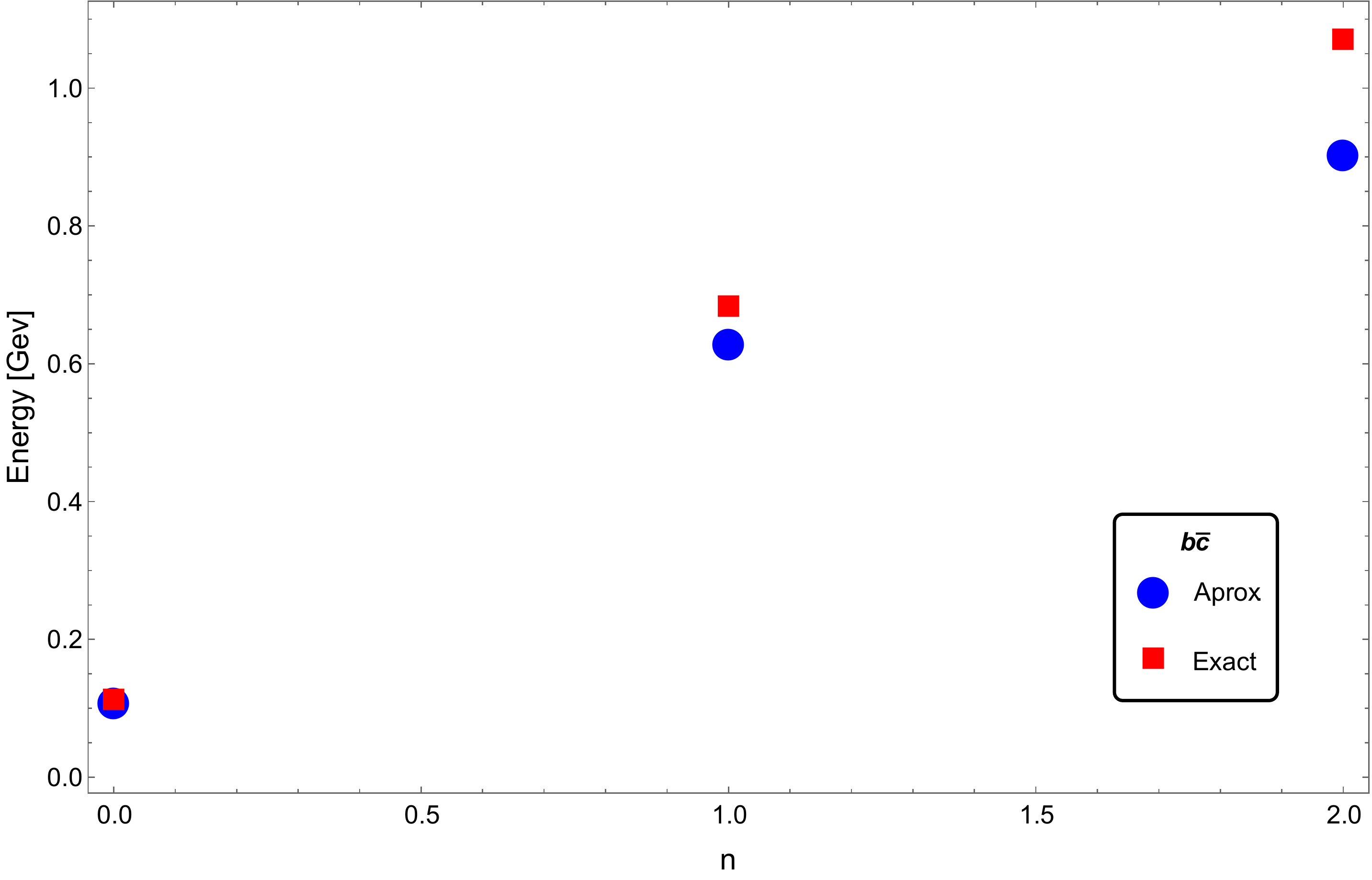}
  \end{tabular}
\caption{Plots shown a comparison in energies for first three energy values for heavy quarkonium $c\bar{c}$, $b\bar{b}$ y $b\bar{c}$. Square correspond to "exact values " calculated with mathschroe.nb with step $h=0.00001$, and circles represent values calculated in this paper.}
\end{figure*}
\end{center}

Now we calculate WFO for the ground state. In this case we consider two approaches that are equivalent when you work with the exact solutions, but they give different values if you use approximated wave functions. Here we use $R(r)=\frac{U(r)}{r}$.

The first approach considered for calculations of WFO, and from now on call ``Method 1'', is based in a well known expresion that relate WFO with expectation values for first derivative of the potential,
\[
\left\vert \Psi (0)\right\vert ^{2}=\frac{\mu }{2\pi }\left\langle \frac{%
dV(r)}{dr}\right\rangle 
\]
for S states
\[
\Psi (r)=R(r)Y(\theta ,\varphi )=\frac{1}{\sqrt{4\pi }}R(r) 
\]
so
\[
\left\vert R(0)\right\vert ^{2}=2\mu \left\langle \frac{dV(r)}{dr}%
\right\rangle.
\]

\begin{table*}
\begin{center}
\caption{Comparison of WFO for first three energy values for heavy quarkonium $c\bar{c}$, $b\bar{b}$ y $b\bar{c}$. We show values calculated with numerical solution, with method 1 and 2.} 
\vspace*{.25cm}

\def\arraystretch{1.5}
\begin{tabular}{|c|c|c|c|c|c|c|c|c|c|}
\hline
$WFO$ & \multicolumn{3}{|c|}{$c\overline{c}$} & \multicolumn{3}{|c|}{$b%
\overline{b}$} & \multicolumn{3}{|c|}{$b\overline{c}$} \\ \hline
$\left\vert R\left( 0\right) \right\vert ^{2}$ & $Exact$ & $Method~1$ & $Method~2$ & $Exact$ & $Method~1$ & $Method~2$ & $Exact$ & $Method~1$ & $Method~2$ \\ \hline
$1s$ & $1.4591$ & $1.4384$ & $1.2897$ & $14.1294$ & $13.9824$ & $13.0031$ & $%
3.1950$ & $3.1486$ & $2.8380$ \\ \hline
$2s$ & $0.9304$ & $0.8160$ & $0.6631$ & $5.7033$ & $4.1764$ & $3.3871$ & $%
1.7712$ & $1.4551$ & $1.1601$ \\ \hline
$3s$ & $0.7936$ & $0.6781$ & $0.5186$ & $4.2917$ & $2.6210$ & $1.8017$ & $%
1.4509$ & $1.1171$ & $0.8048$ \\ \hline
\end{tabular}

\end{center}
\end{table*}

The second approach is simply to take $r \rightarrow 0$ in the wave function. This is what we called ``method 2" in the next lines. According to this, WFO for ground state can be found directly from $R(r)=\frac{U(r)}{r}$ and (\ref{FnEnsayo}) (with $\gamma = 1$ for ground state), and we get

\[
\left\vert R(0)\right\vert ^{2}=N^{2}=\frac{\left( 2a_{0}\right) ^{\frac{%
1+2\gamma }{b_{0}}}b_{0}}{\Gamma \left( \frac{1+2\gamma }{b_{0}}\right) } 
\]

Using the variational method to get approximations to the excited states, it is not so simple, because you need to be sure that yours trial eigenfunctions are orthogonal. For this reason we consider to solve the problem of the ground states for the supersymmetric partner. Let us consider how to use the SUSY QM and the variational method to get solutions to states 2S.

Previously we use (\ref{FnEnsayo}) with $\gamma = 1$ to get solutions for the ground state, so using the usual variational method we can get energy values for the states 1S, and we can fix parameter in the wave functions, and we put an index ``0" to remember us that correspond to parameters associated to the ground state. With this trial wave function we obtain the superpotential
\[
W_{21}(r)=-\frac{1}{\sqrt{2\mu }}\frac{U^{\prime }(r)}{U(r)}=%
\frac{-1+a_{0}b_{0}r^{b_{0}}}{\sqrt{2\mu }r}.
\]
here index 21 in $W$ remember us that starting from solutions of potential $V_{1}$ (Cornell in this paper) we can built a potential $V_{2}$ (an approximated supersymmetric partner for Cornell potential)
\[
V_{2}(r)=\left[ W_{21}(r)\right] ^{2}+\frac{1}{\sqrt{2\mu }}%
\left( \frac{dW_{21}(r)}{dr}\right) 
\]%
\[
V_{2}(r)=\left( \frac{%
2+a_{0}b_{0}r^{b_{0}}(-3+a_{0}b_{0}+a_{0}b_{0}r^{b_{0}})}{2\mu r^{2}}\right) 
\]

\begin{center}
\begin{figure*}[ht]
  \begin{tabular}{c c c}
    \includegraphics[width=2.3 in]{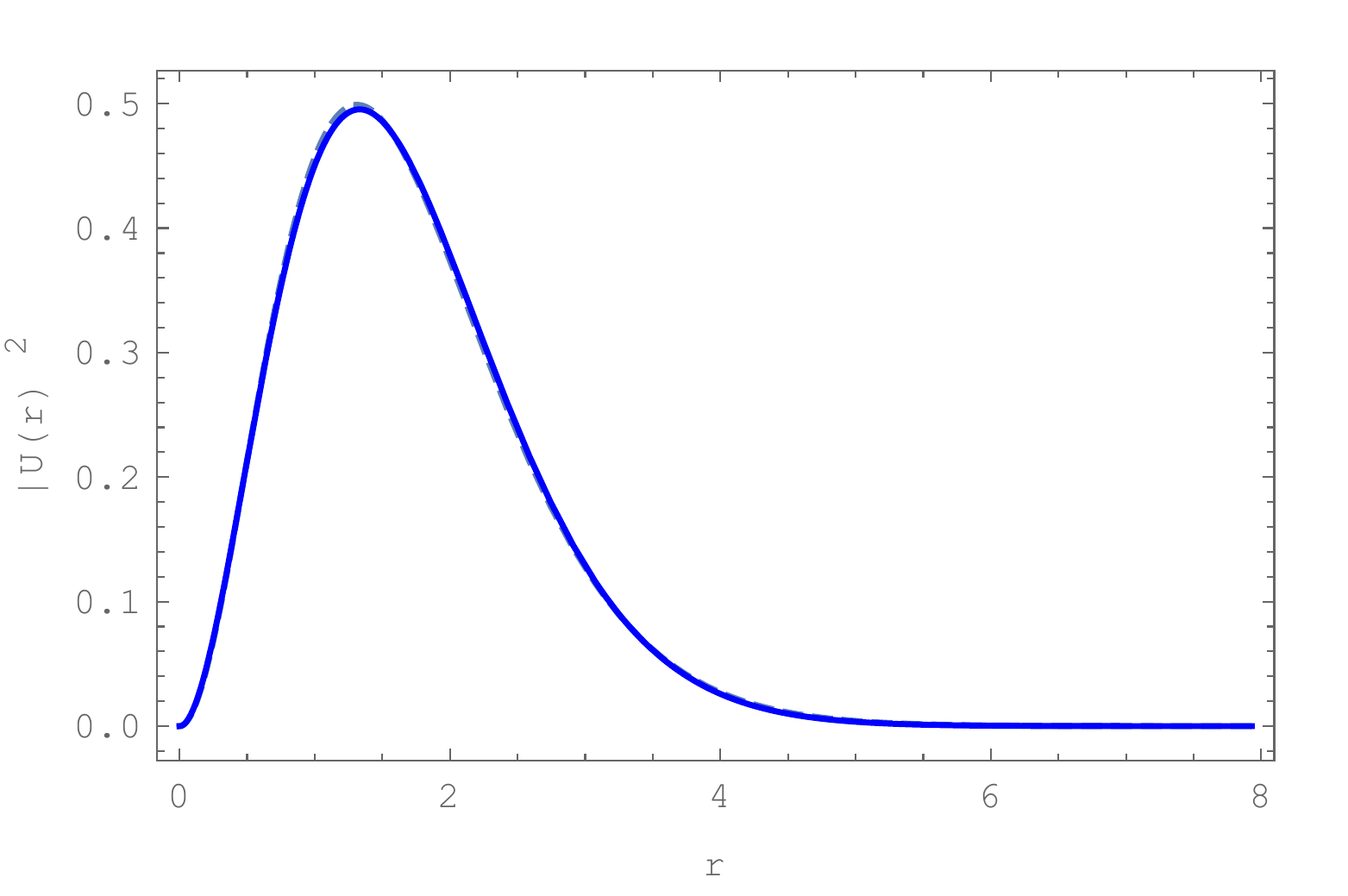}
    \includegraphics[width=2.3 in]{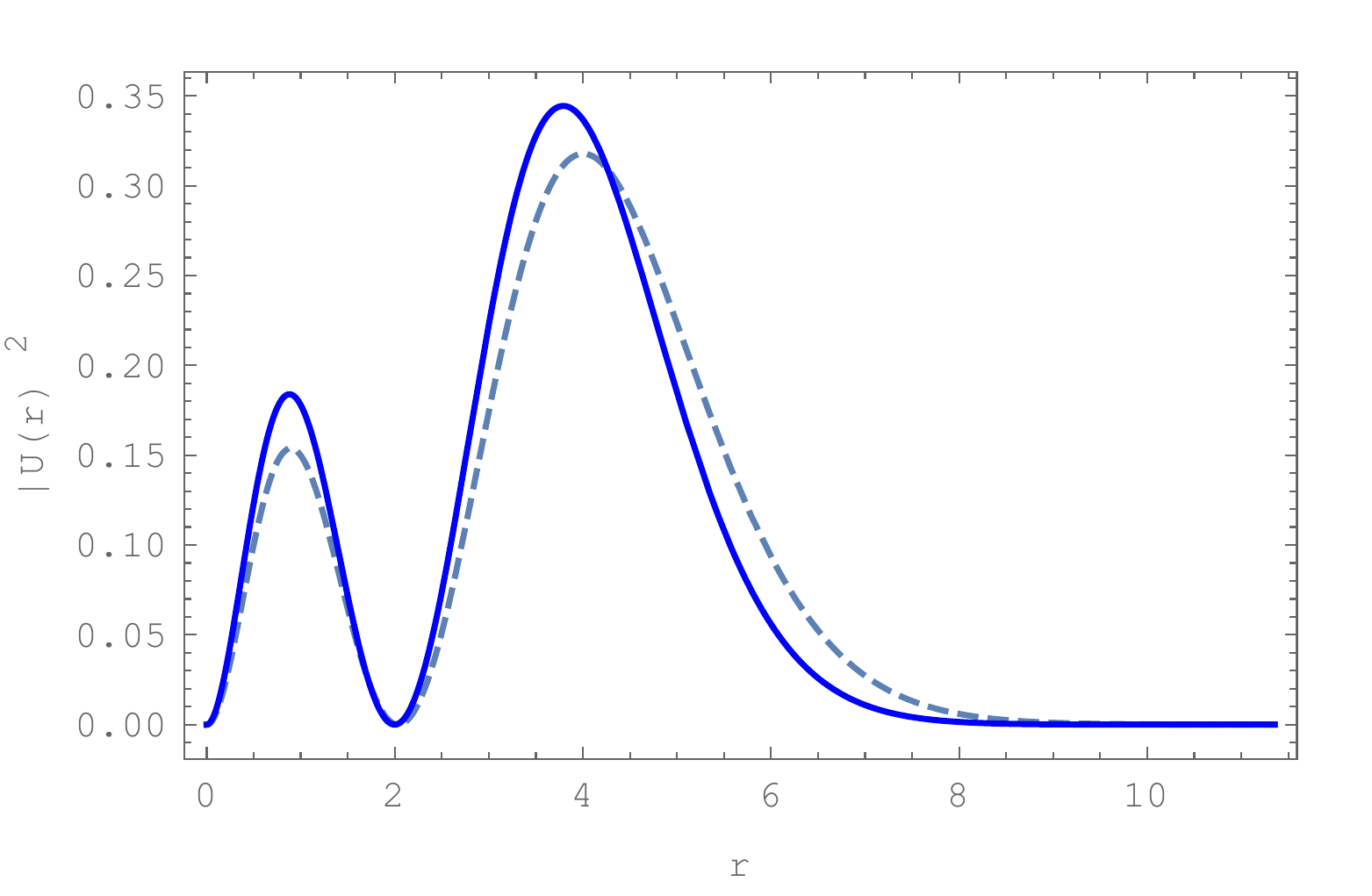}
    \includegraphics[width=2.3 in]{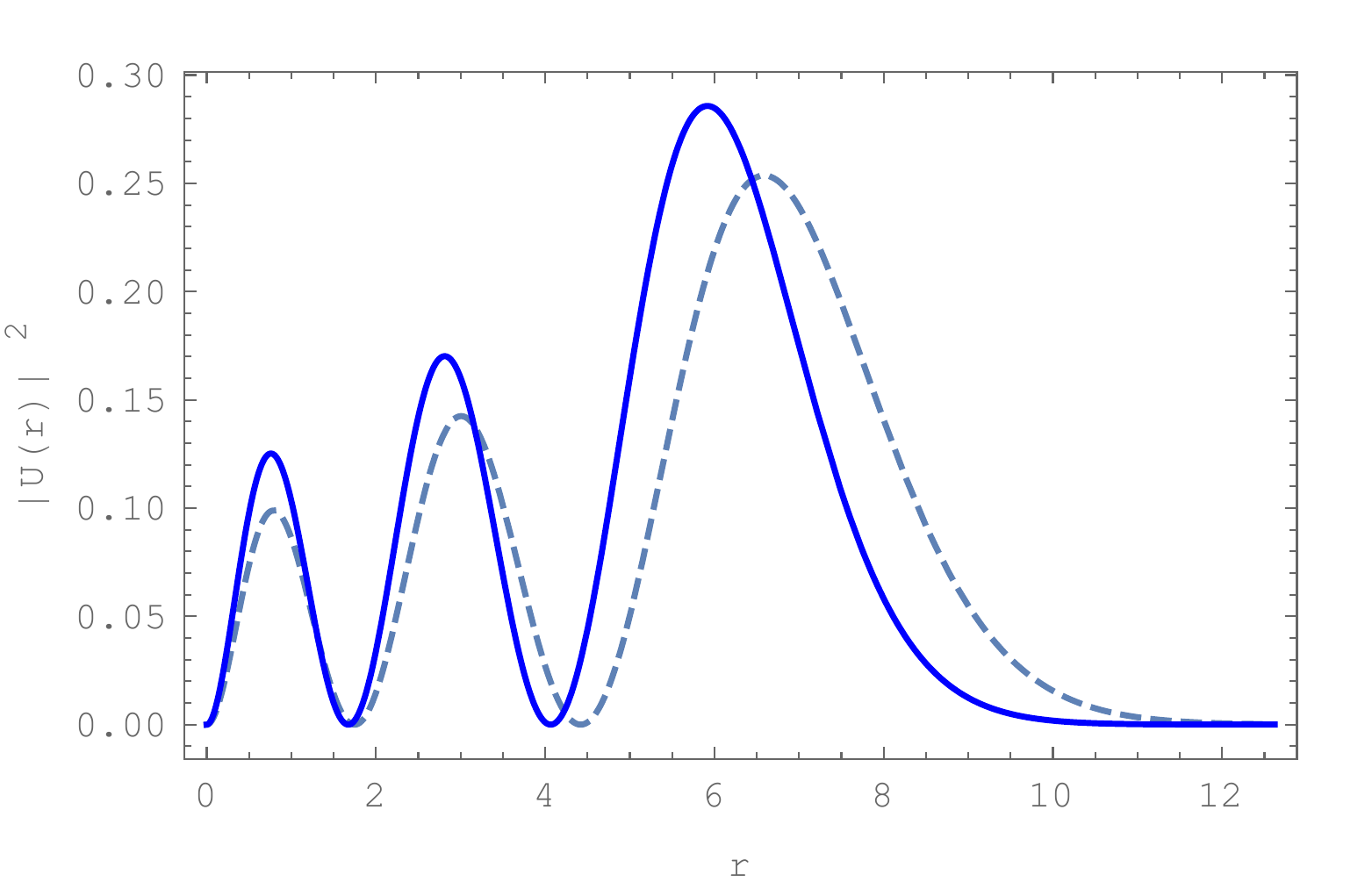}\\
    \includegraphics[width=2.3 in]{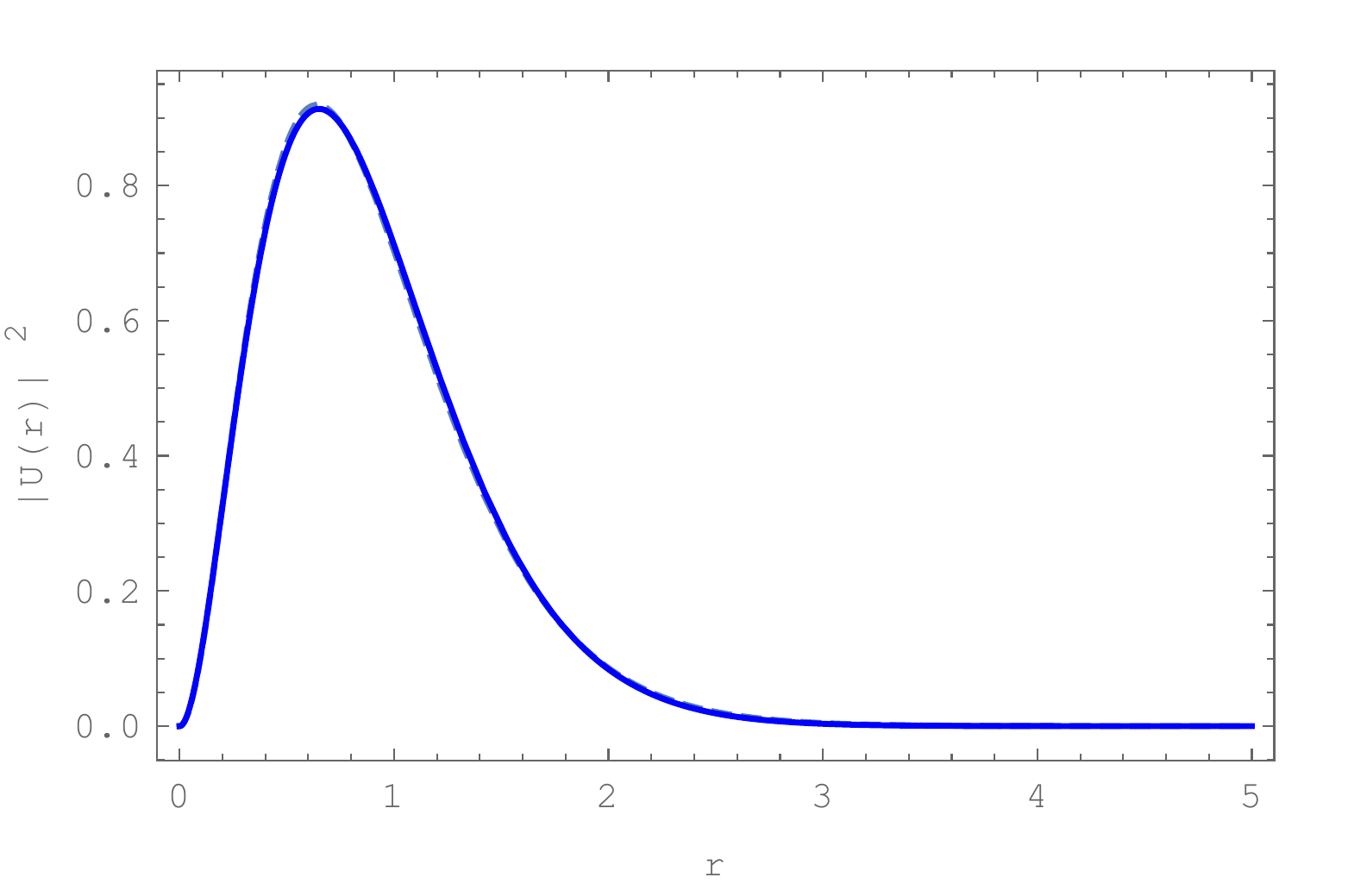}
    \includegraphics[width=2.3 in]{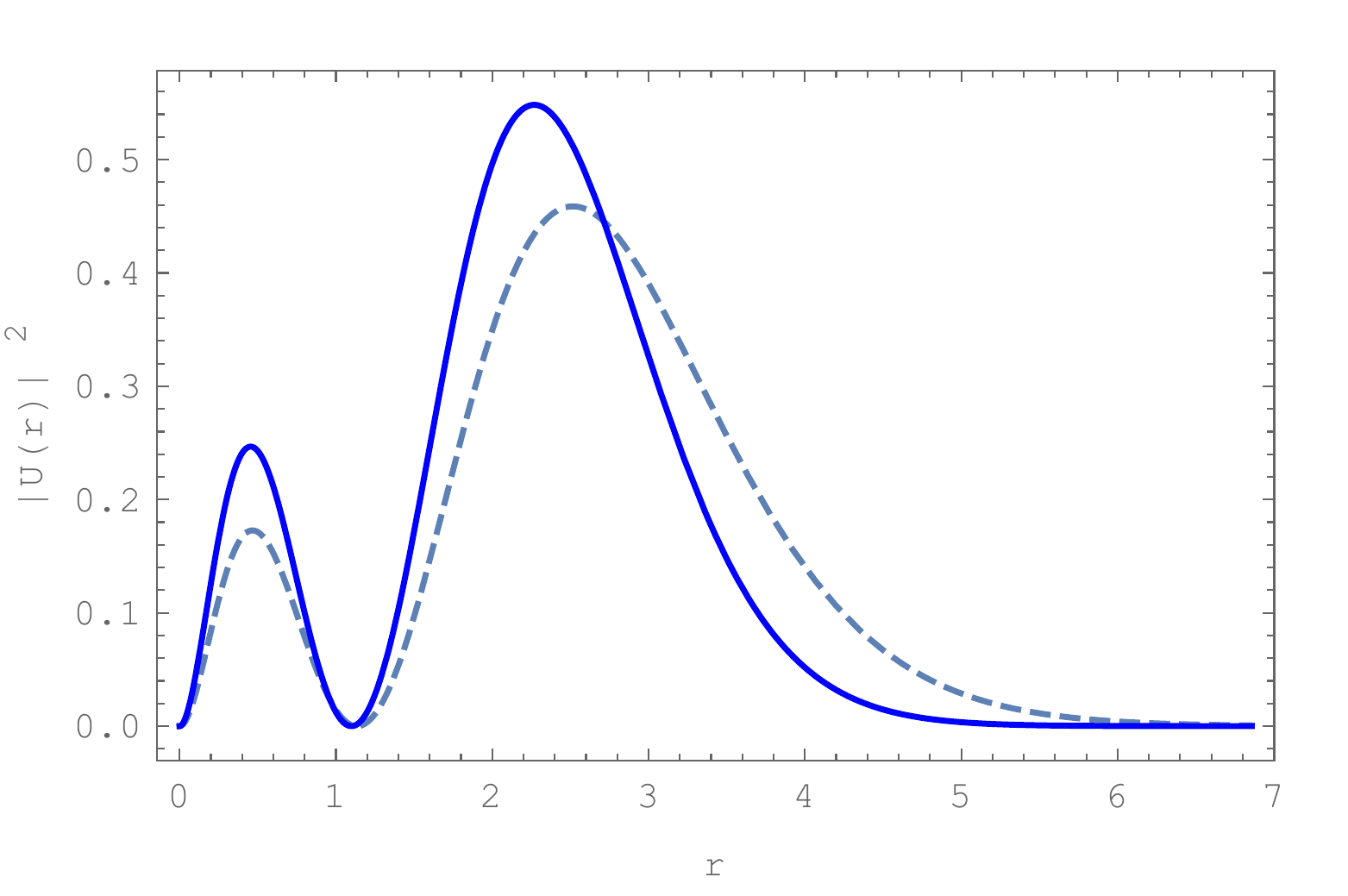}
    \includegraphics[width=2.3 in]{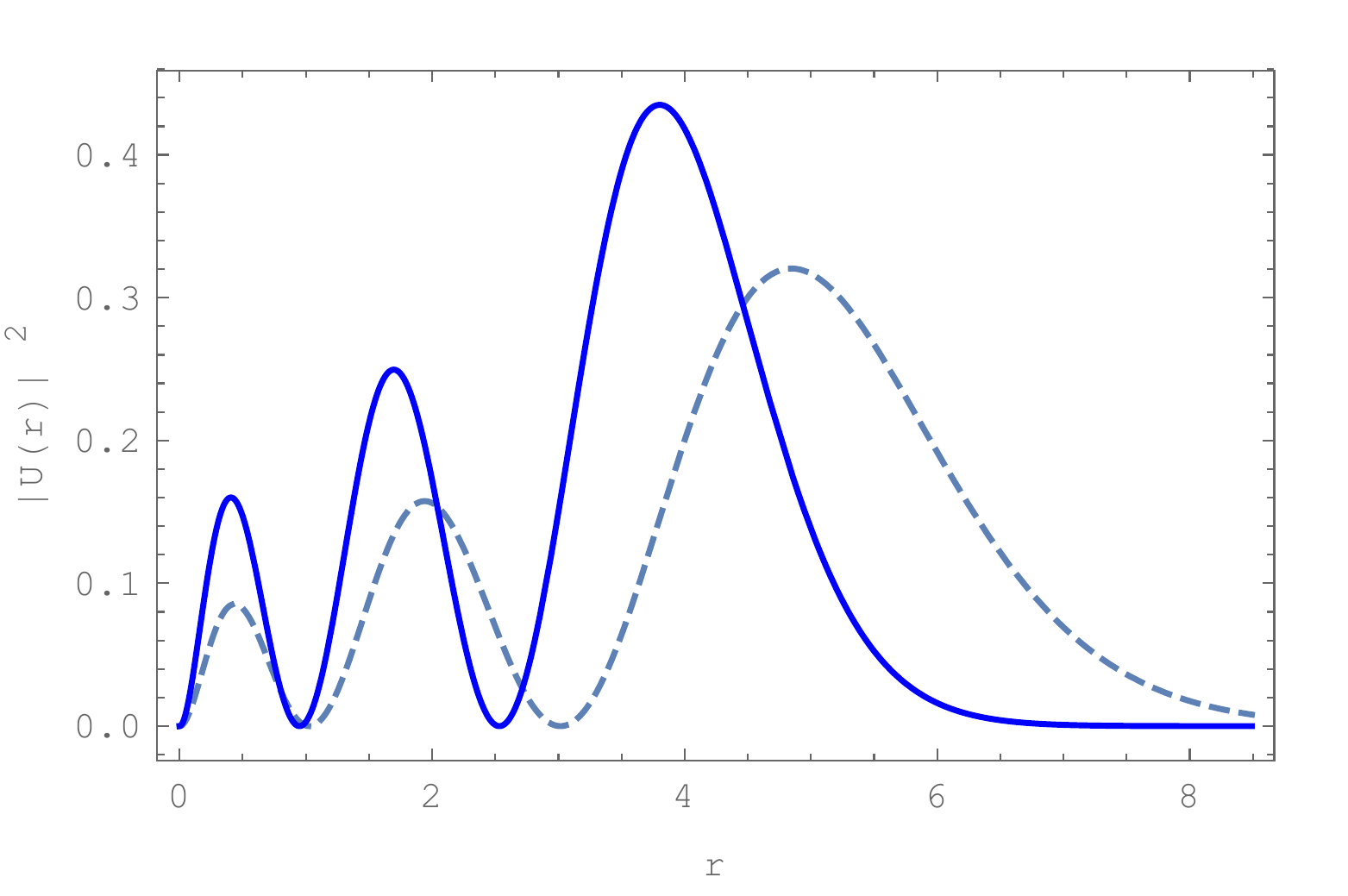}\\
    \includegraphics[width=2.3 in]{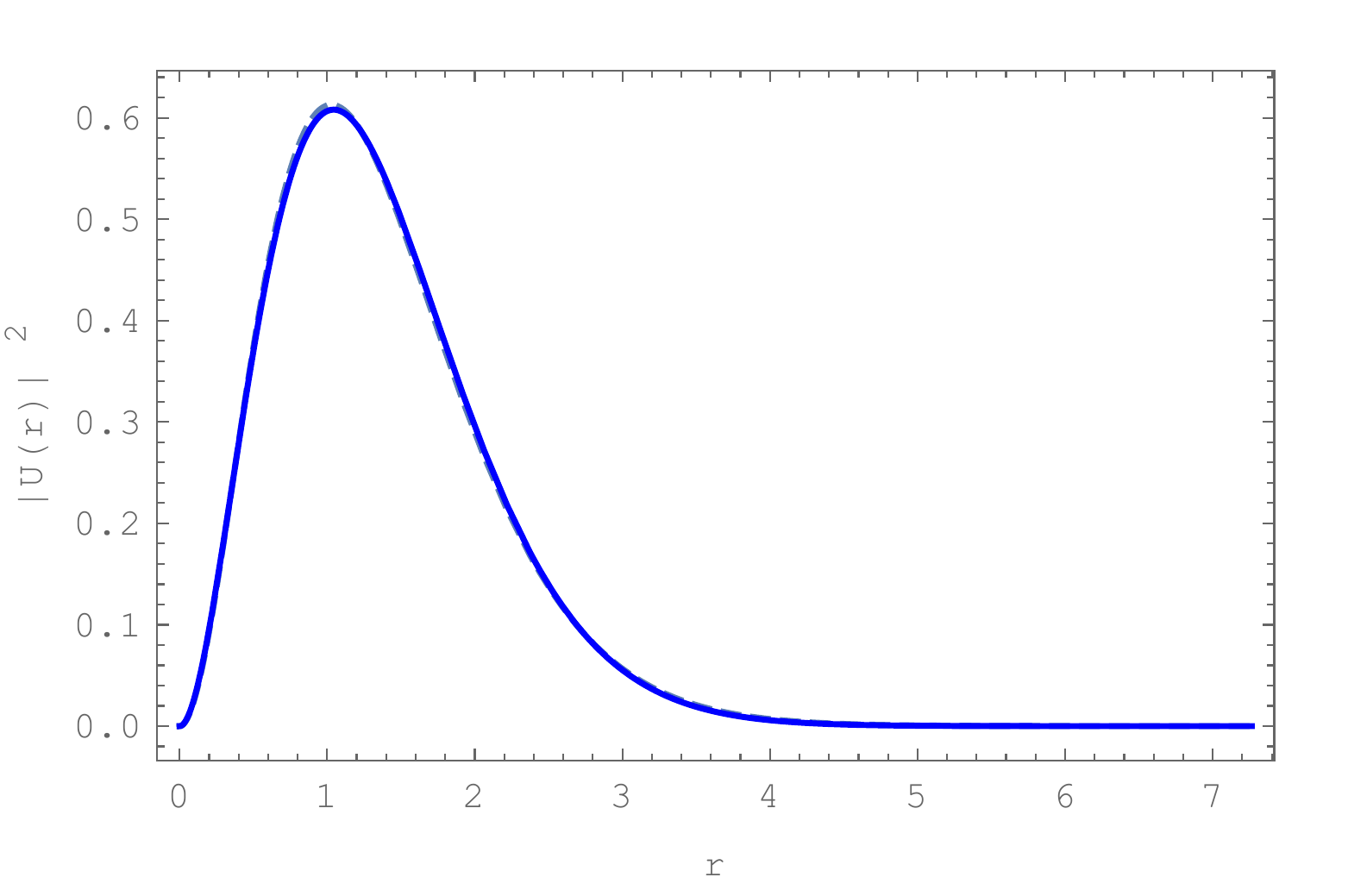}
    \includegraphics[width=2.3 in]{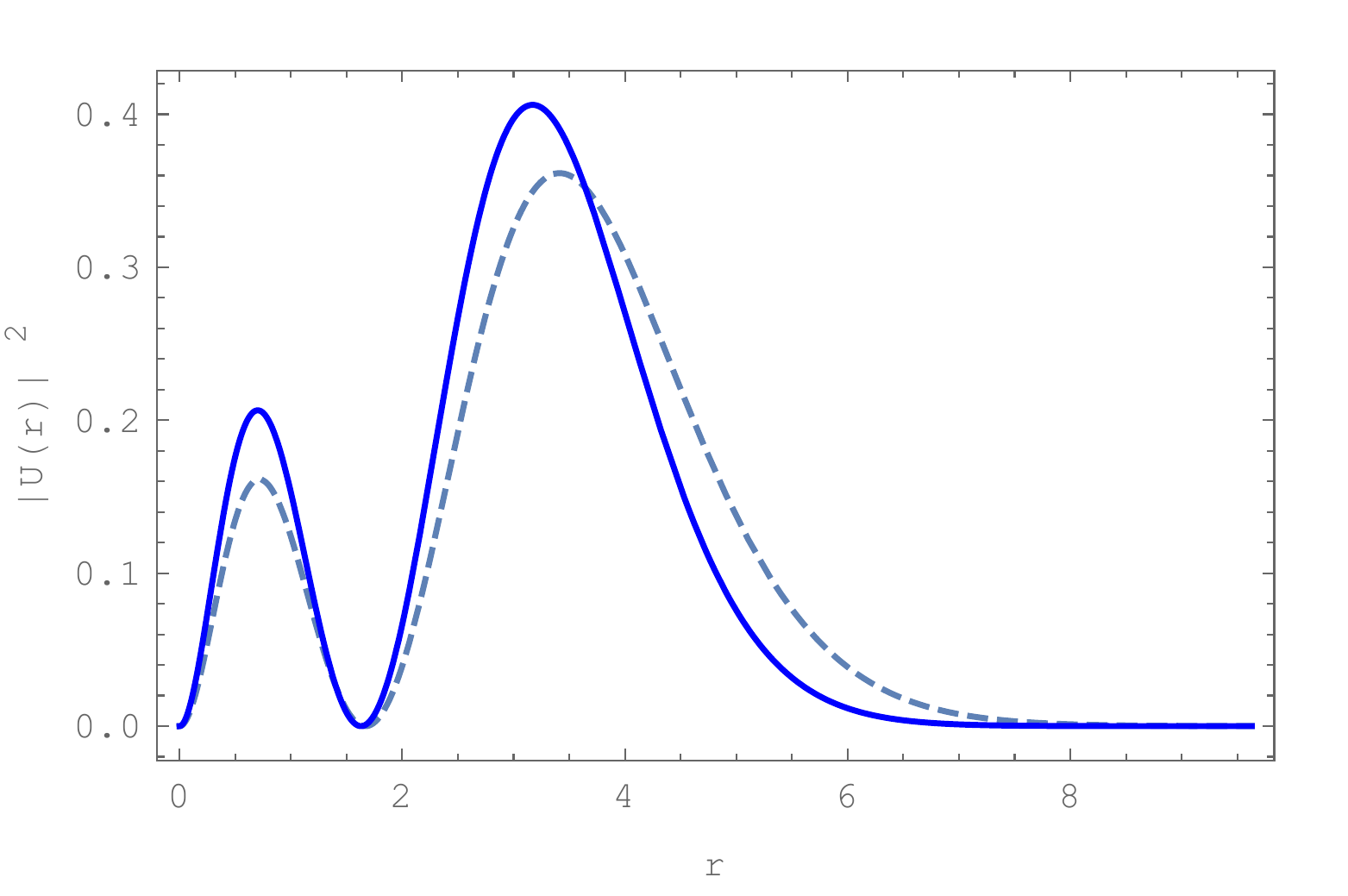}
    \includegraphics[width=2.3 in]{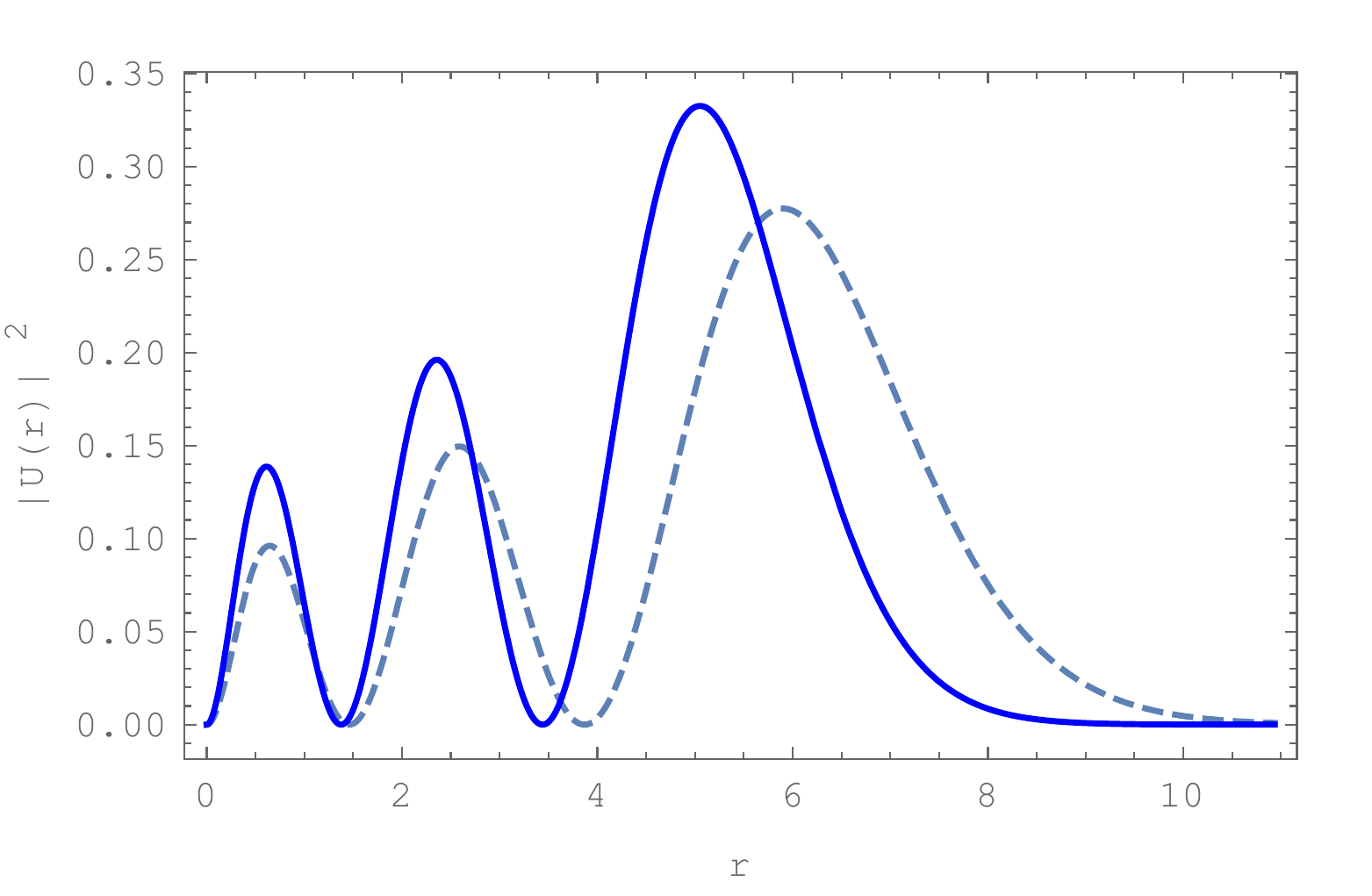}
  \end{tabular}
\caption{Radial probability density functions. The continous line is numerical and dashed line correspond to method used in this paper. First column correspond to ground states, second column is for first excited state and third column is for second excited stated. The upper row is for $c\bar{c}$, middle row is for $b\bar{b}$ and lower row is for $b\bar{c}$.}
\end{figure*}
\end{center}


Now, using the variational method we get an approximated value for ground state energy of potential $V_{2}$. In this case the trial wave function has the shape (\ref{FnEnsayo}) with $\gamma = 2$, and the energy is related to the first excited state of $V_{1}$.

Using the variational method we get the expectation value for energy that depend on $a_{0}$, $b_{0}$ (fixed in previous steps when we calculate the ground state of $V_{1}$) and $a$ and $b$, that must be fixed once we minimize this expectation value. We add an index ``1'' in parameters $a$ and $b$.

Notice that, in our discussion for SUSY QM we consider a ground state with eigenvalue equal to zero, so the ground state of energy for $V_{2}$ represent a $\Delta E_{2}$, so the energy for the first excited state is
\[
E_{1}=E_{0}+\Delta E_{2}
\]

Now we can continue. As we had an approximated solution for $V_{2}$, we can built a $W_{32}$ and from it to get his supersymmetric partner $V_{3}$. If we found the ground state of this new potential using the variational method, we can found the energy of the second excited state by
\[
E_{2}=E_{1}+\Delta E_{3},
\]
and so on. Table 1 and FIG. 2 show the energy values calculated with the method used in this paper, and we compare it with exact numerical solution that we get using a MATHEMATICA program called mathschroe.nb \cite{Lucha:1998xc}.

Let us consider the wave function for the excited levels of Cornell potential. If we have a solution for the ground state of the potential $V_{2}$, that we call for example $\psi_{0}^{(2)}$, and as we know $W_{21}$ (we used to built $V_{2}$), it is possible to get a wave function for the first excited state of $V_{1}$ using
\[
\psi _{1}^{(1)}\sim A_{21}^{\dagger }\psi _{0}^{(2)},
\]
In principle, for exact normalized solutions, this transformation give the right normalization, but as we are working with approximated solutions we prefer normalize each wave function at the end, for this reason in previous expression we use symbol "$\sim$".

$A_{21}^{\dagger }$ transform solution $\psi _{0}^{(2)}$ for the ground state of $V_{2}$ in a solution for the first excited state of $V_{1}$, so it gave us the wave function of the first excited state of Cornell potential.

\[
\psi _{1}^{(1)}\sim \left( \frac{-1}{\sqrt{2\mu }}\partial r+W_{21}(r)\right) \psi _{0}^{(2)}
\]

In a similar way, we can build the wave function for the second excited state of Cornell potential, if we start from the ground state of solution for $V_{3}$
\[
\psi _{2}^{(1)}\sim A_{21}^{\dagger }A_{32}^{\dagger }\psi _{0}^{(3)}
\]
\[
\psi _{2}^{(1)}\sim \left( \frac{-1}{\sqrt{2\mu }}\partial r+W_{21}(r)\right) \left( \frac{-1}{\sqrt{2\mu }}\partial r+W_{32}(r)\right) \psi _{0}^{(3)}
\]

In FIG. 3 we show the radial density of probabilities calculated with this method, and comparing it with the result obtained numerically using mathschroe.nb. We use the wave functions to calculate WFO, and table III and FIG 4 show a summary of our results obtained using ``Method 1'' and ``Method 2''.

\begin{center}
\begin{figure*}[t]
  \begin{tabular}{c}
    \includegraphics[width=2.3 in]{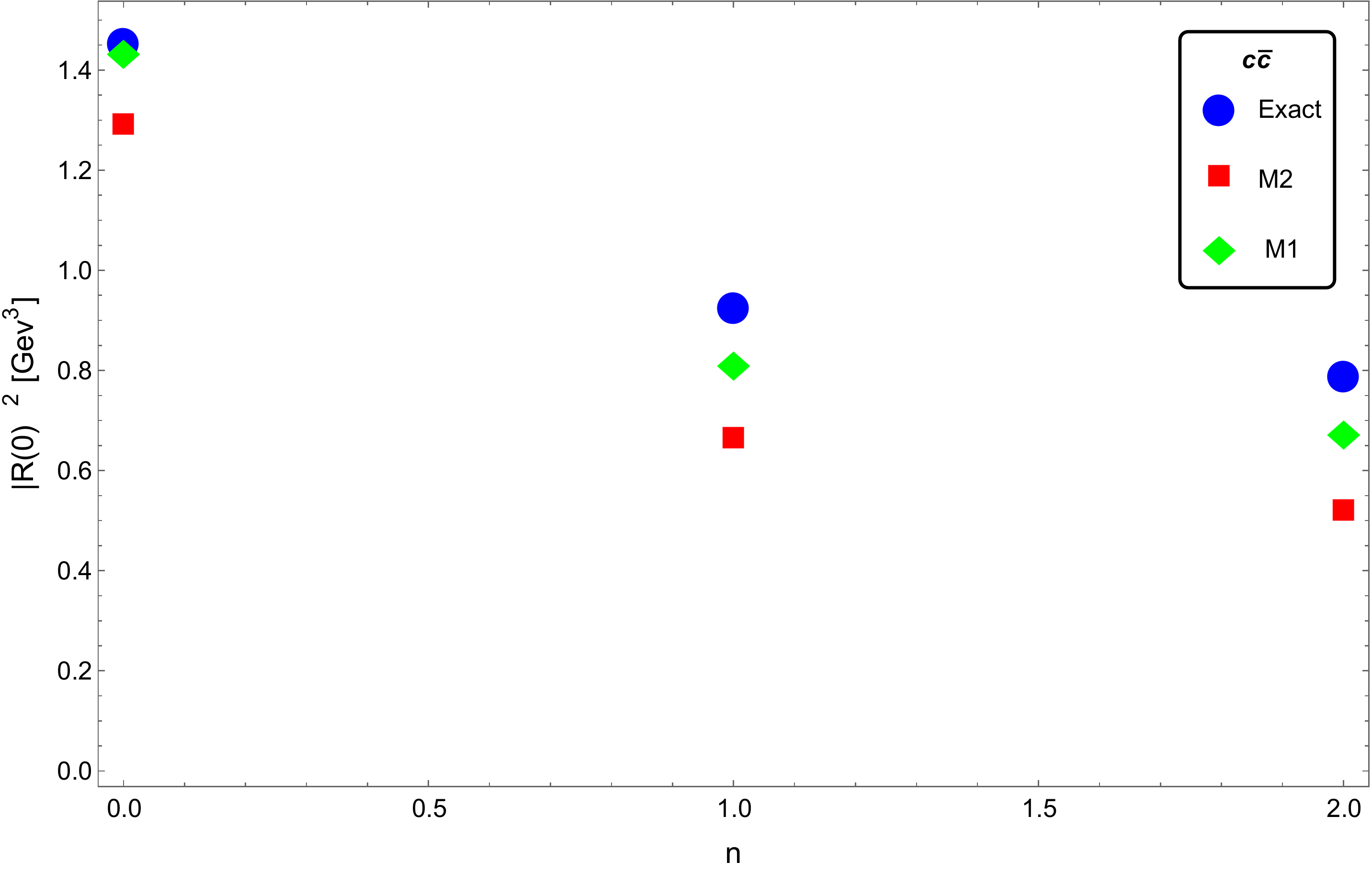}
    \includegraphics[width=2.3 in]{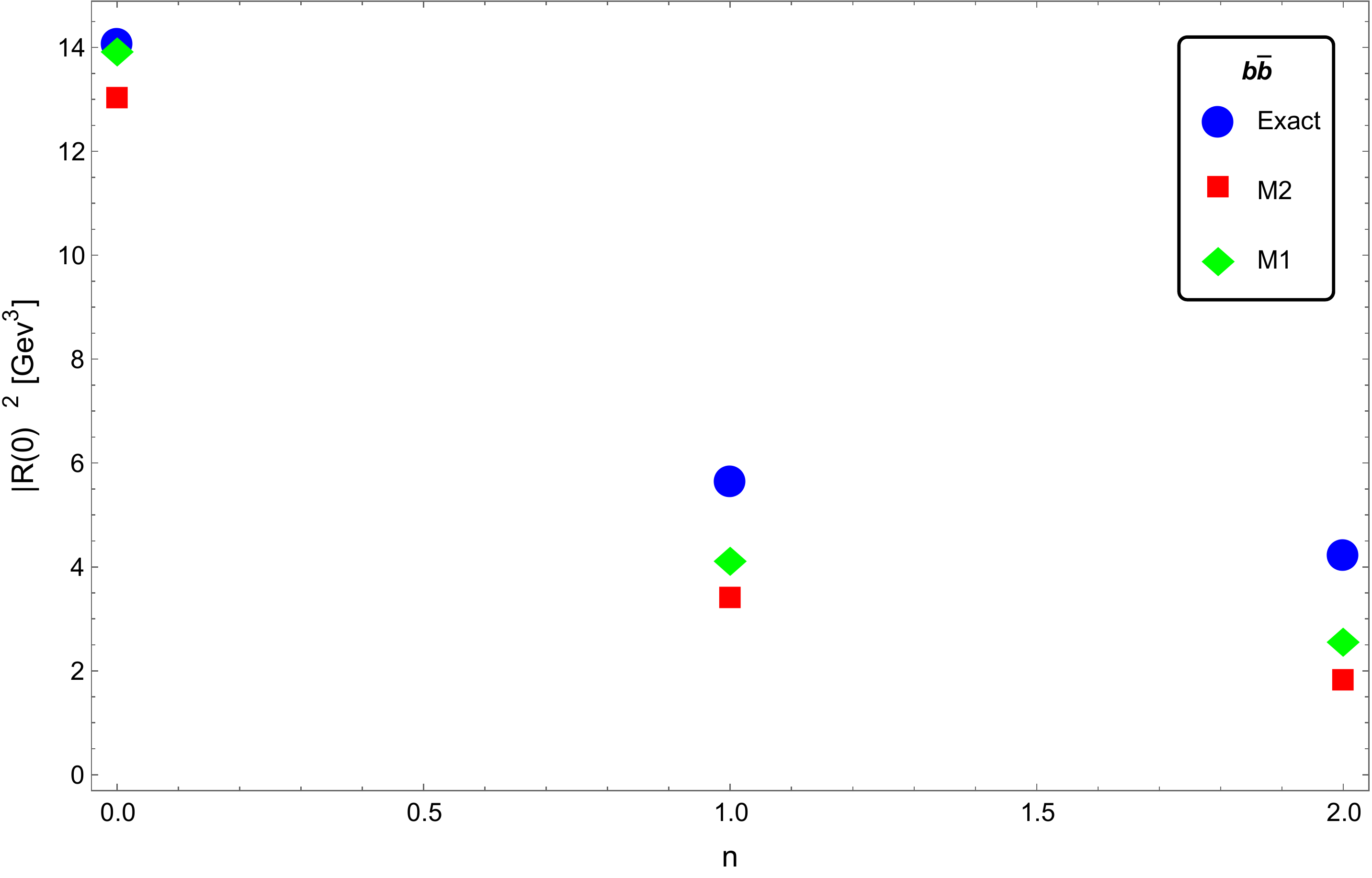}
    \includegraphics[width=2.3 in]{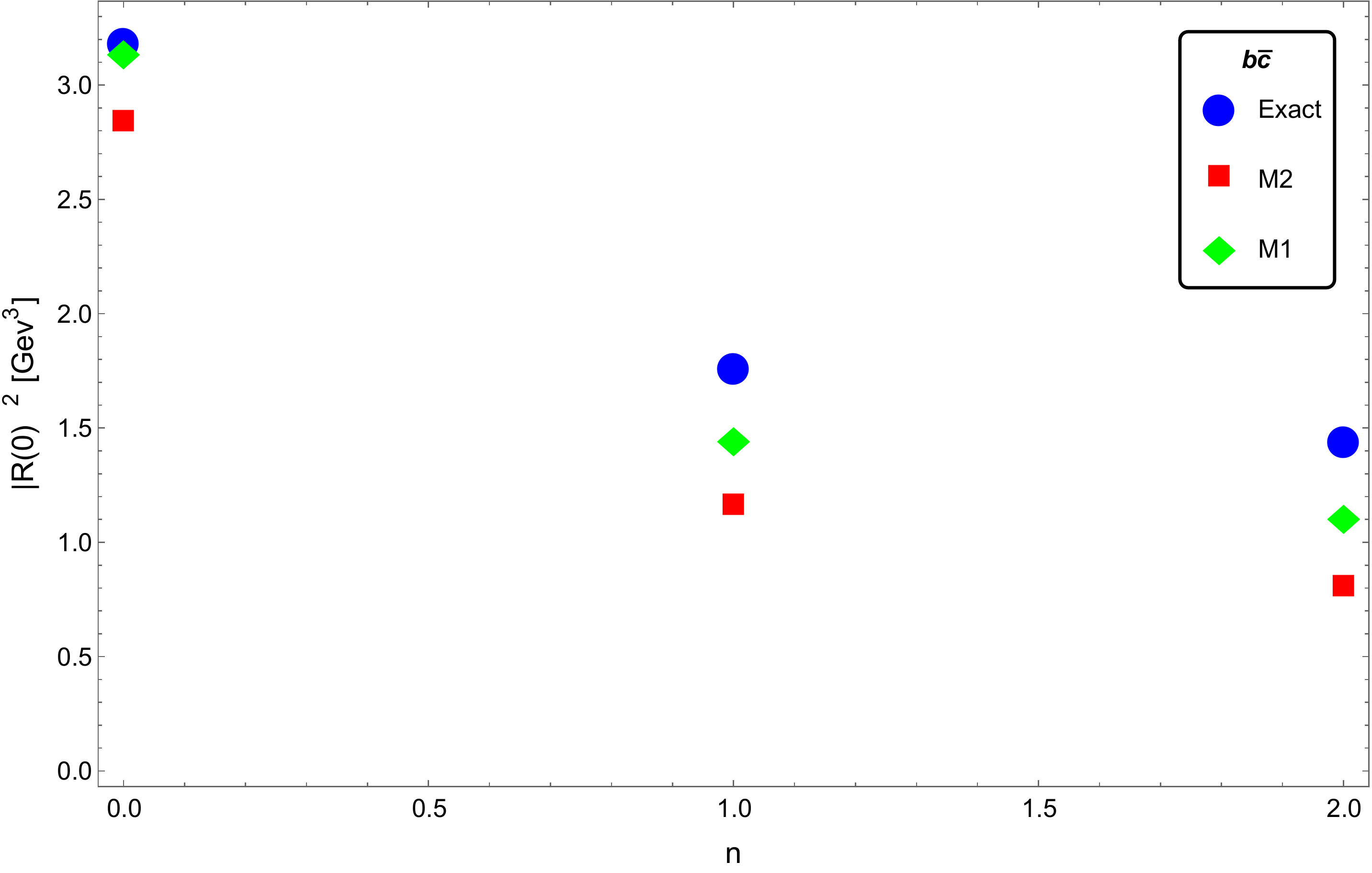}
  \end{tabular}
\caption{Plots shown a comparison in WFO for first three energy values for heavy quarkonium $c\bar{c}$, $b\bar{b}$ y $b\bar{c}$. Circles correspond to "exact values", diamond were calculated with method 1 and squared are values calculated using method 2.}
\end{figure*}
\end{center}

\section{Conclusion and discussion}

We use a procedure to solve an approximate way the Schr\"odinger equation with Cornell potential using a variational method and SUSY QM. This is phenomenologically interesting, because Cornell potential can describe some properties of heavy quarkonium, for this reason is useful to have analytical wave functions, as the provided in this paper

The results in TABLE I and FIG. 2 show the values for the energies of the first three excited states, that are in good agreement with the exact computation, specially for the ground state and first excited states. This situation remain once we compare the wave functions as you can see in FIG. 3, that show radial probability density functions. Numerical and approximate wave functions are almost the same for the ground state and very close for the first excited state, but when we consider higher radial excitations both wave functions are different.

We also calculate WFO, for this we consider two methods that are equivalent when you work with exact wave functions, but as you can see in TABLE II, they give different results if you use approximated wave functions. The ``Method 1", based in calculations of the expectation values of first derivative of potential give a better result than ``Method 2".

Our results suggest that the exact method discussed can give results in agreement with numerical exact solutions for lower states, and disagree for higher excited levels. This is not surprising, because we solve the Schr\"odinger equation in an approximated way with an approximated potential. So even if we start with a good trial wave function for the ground state, for higher radial excitations the results will be in disagreement with the numerical solutions, but for lower states this approach provide good results. If we use trial wave functions with several parameters it could be possible to improve our results, but as this method work for lower states, maybe could be useful to use it as a complement with a method that works for higher excitations as WKB.

As this approach give us analytical expressions for wave functions close to numerical solutions and they are orthogonal, so this can be useful to do calculations of other heavy quarkonium properties.

\begin{acknowledgments}

This work was supported by FONDECYT (Chile) under Grant No. 1141280 and by CONICYT (Chile) under Grant No. 7912010025.

\end{acknowledgments}

\end{document}